\documentclass[11pt]{article}
\usepackage{amsmath}
\usepackage{amssymb}
\usepackage{graphicx}
\usepackage{color}
\usepackage{cancel}
\usepackage{fancybox}

\numberwithin{equation}{section}

\begin{document}
\begin{titlepage}
\begin{flushright}
HRI-P-08-11-004 \\
\end{flushright}

\begin{center}
\vspace*{15mm}
{\Large \bf Hybrid textures in minimal seesaw mass matrices
}
\vspace{.5in}

Srubabati Goswami$^{1,2}$\footnote{sruba@prl.res.in}, 
~Subrata Khan$^{1}$\footnote{subrata@prl.res.in},
~Atsushi Watanabe$^2$\footnote{watanabe@mri.ernet.in}
\vskip 0.5cm
$^1${\small {\it Physical Research Laboratory, Navrangpura, 
Ahmedabad -380009, India}},\\
$^2${\small {\it Harish-Chandra Research Institute, Chhatnag Road, Jhunsi,
Allahabad -211019, India}}\\

\vskip 1in

\end{center}

\begin{abstract}\noindent%
In the context of minimal seesaw framework,  
we study the implications of Dirac and Majorana 
mass matrices in which two rigid properties coexist, namely, 
equalities among mass matrix elements
and texture zeros.
In the first part of the study, we discuss general possibilities of
the Dirac and Majorana mass matrices for neutrinos with
such hybrid structures. 
We then classify the mass matrices into realistic textures
which are compatible with global neutrino oscillation data 
and 
unrealistic ones which do not comply with the data. 
Among the large number of general possibilities,
we find that only 6 patterns are consistent 
with the observations at the level of the most minimal number of 
free parameters.
These solutions have only 2 adjustable parameters, so that
all the mixing angles can be described in terms of the two mass differences
or pure numbers.
We analyze these textures in detail and discuss their impacts for 
future neutrino experiments and for leptogenesis. 
\end{abstract} 

\end{titlepage}
\newpage
\section{Introduction}
The origin of the generation structure realized in nature is an
engrossing subject which has been discussed for a long time, 
but still remains veiled.
Within the standard electroweak theory, masses and mixing angles
of fermions originate from the Yukawa interactions with the Higgs boson, 
responsible for the electroweak symmetry breaking.
While, the relative strength of the gauge interactions for various
fermion species are controlled by gauge invariance, 
Yukawa couplings are not governed by any principle.
They bring a multitude of free parameters into the theory and even have
ambiguities in reconstructing their values from experiments. 
A viable approach is, therefore, to search for mass matrices
which are taking suggestive forms in light of model building 
ingredients, such as symmetry amongst generations.

A direct scheme in this spirit is ``texture zero'' in the
mass matrices of fermions.
In this framework, it is assumed that the mass matrices have several
elements which are anomalously small compared to the others.
Initially this approach was studied in the quark sector \cite{quarkM} 
and it was found that the existing relations among masses and mixings 
of quarks can be explained by the vanishing matrix elements.
The available textures presented in the literature provide foundations
of model building and insights into the generation puzzle.

As for the lepton sector, recent progress in neutrino physics
makes it possible to discuss feasible forms for mass matrices.  
In \cite{Frampton:2002yf} and many subsequent papers, 
the texture zeros in lepton sector have been discussed in the context of 
various 
types of mass  matrices 
such as the Majorana mass matrix of the left-handed neutrinos \cite{mnuzero},
both the charged lepton and neutrino 
mass matrices \cite{chargedleptonTEX},
the Dirac and the Majorana mass matrices in the seesaw
mechanism \cite{seesawTEX}
and amalgamated Yukawa couplings in grand unification
\cite{gut}.
All these results also provide useful information to infer the structure
lying behind the Yukawa interactions.

It is to be noted, that,   
for the mass matrices in the  lepton sector, the situation
is different from that of the quark sector because of
the weak hierarchy of neutrino mass spectrum (for normal hierarchy)
and two large mixing angles.
In addition, the neutrino spectrum can also have inverted hierarchy 
and can even be quasi-degenerate which has no analogue in  the 
quark sector. 
One of the most 
distinct features amongst observations in the neutrino sector  
is the small 1-3 angle against  possible maximal 2-3  mixing.  
This is the origin of the  $\mu$-$\tau$ symmetric nature of the 
neutrino mass matrix  \cite{mutau}.  
In the generation basis where the charged-lepton mass matrix is
diagonal, the exact $\mu$-$\tau$ symmetric limit  means that there are simple 
equalities among the matrix elements of the neutrino mass matrix.  
Moreover, the $\mu$-$\tau$ symmetric neutrino mass matrix contains
the tri-bimaximal mixing \cite{tribi} as a special case, which is obtained by
requiring further special relations among the matrix elements.
Thus, the equality between the matrix elements might play an important 
role in understanding the properties of neutrinos
and become a simple and direct approach to search for 
``suggestive forms'' of the mass textures.

The equalities among left-handed Majorana mass matrix elements  
were discussed in \cite{frigerio-smirnov} in a bottom-up way. 
In \cite{Kaneko:2005yz}, hybrid structures, with 
the coexistence of equalities among
matrix elements and texture zeros for the left-handed 
Majorana mass matrix were studied. 
In this paper we follow the approach of \cite{Kaneko:2005yz}, 
but consider the equalities and  zeros in the 
Dirac and  right-handed
Majorana mass matrices in the seesaw mechanism. 
We perform a thorough classification of such hybrid textures and 
identify the left-handed Majorana mass matrices that can be  
reconciled with the inputs obtained from neutrino oscillation data. 
In particular, we consider the case where only 
two right-handed neutrinos take part in the seesaw mechanism 
\cite{2nuR}. Texture analysis in the context of such a minimal 
seesaw scheme has been accomplished in \cite{2nuRTEX,2nuRus}.
An advantage of this choice is that there are less number of  free parameters.
Thus the mass matrices get simple forms and rich predictions compared
to the standard three heavy neutrino models.
By exhausting all possibilities, we find a novel class of textures
which have only two adjustable parameters to fit the low-energy data.
We discuss the possibility of having leptogenesis in these textures 
and explore the connection between high and low energy CP violation 
if any. 

The layout of the paper goes as follows.
In Section 2, we present the formulations which are needed for  
the subsequent discussions.
In Section 3, we study equalities amongst mass matrix elements and
enumerate general possible forms of the mass matrices with equality
relations.
In Section 4, we perform a combined analysis of equalities and
texture zeros and discuss the viable minimal textures
which are compatible with the current neutrino data.
In Section 5, we discuss CP violation, paying particular
attention to possible connections between leptogenesis and
CP violation in neutrino oscillation. 
In Section 6, we briefly comment on the renormalization group
effects for the mass matrices.
Section 7 is devoted to conclusion and summary.

\section{Formulation and the oscillation parameters}
We assume that the tiny neutrino masses are
generated through type-I seesaw mechanism \cite{seesaw1, seesaw2}
by the suppression effect of the large mass scale of
the right-handed neutrinos.
By integrating out the heavy right-handed neutrinos,
the Majorana mass matrix of the left-handed neutrinos
is obtained as,  
\begin{eqnarray}
\mathcal{M} = - m_D^{\rm T} M_R^{-1} m_D,
\label{majm}
\end{eqnarray}
where $m_D$ denotes the Dirac mass matrix after the electroweak
symmetry breaking, and $M_R$ is the Majorana mass matrix for
the right-handed neutrinos.
In this paper, we consider the case where we have only two
right-handed neutrinos.
Accordingly, the Dirac mass matrix $m_D$ is a $2 \times 3$ 
rectangular matrix, while $M_R$ is given by a $2 \times 2$ symmetric matrix.
It is noteworthy that 
the two generations of fermions are matched well with
the idea of the doublet (irreducible representation ) 
of discrete groups, which have recently been utilized to address
the observed lepton mixings and masses. 

The mixing matrix in the lepton sector, the so-called 
Pontecorvo-Maki-Nakagawa-Sakata (PMNS) matrix $V$ can be defined 
as the unitary matrix which diagonalizes the
left-handed Majorana mass matrix $\mathcal{M}$
in the generation basis where the charged-lepton mass matrix is diagonal:
\begin{eqnarray}
\mathcal{M}  = V^* 
D
V^\dag ,
\end{eqnarray}
where $D \equiv {\rm diag}(m_1, m_2, m_3)$. 
The neutrino oscillation experiments are capable of determining the squared
mass differences $\Delta m^2_{ij} \equiv m_i^2 - m^2_j$ and
the three mixing angles and one CP violating phase in $V$.
We parameterize the PMNS matrix $V$ as
\begin{eqnarray}
V =
P
\begin{pmatrix}
c_{12} c_{13} & s_{12}c_{13} & s_{13} e^{-i\delta} \\
-s_{12}c_{23}-c_{12}s_{23}s_{13}e^{i\delta}
& c_{12}c_{23}-s_{12}s_{23}s_{13}e^{i\delta} & s_{23}c_{13} \\
s_{12}s_{23}-c_{12}c_{23}s_{13}e^{i\delta}
&-c_{12}s_{23}-s_{12}c_{23}s_{13}e^{i\delta} & c_{23}c_{13}\\
\end{pmatrix}
\begin{pmatrix}
1 & & \\
  &e^{-i\rho}& \\
  & &e^{-i\sigma}\\
\end{pmatrix},
\label{pmns}
\end{eqnarray}
where $c_{ij}$ and $s_{ij}$ stand for $\cos\theta_{ij}$ and $\sin\theta_{ij}$.
The phase $\delta$ represents one phase degree of freedom which is
responsible for CP violation phenomena at low-energy, while
$\rho$ and $\sigma$ are the Majorana phases.
$P$ is a diagonal phase matrix which is to be removed by the 
redefinition of the left-handed fields.
So far, oscillation experiments have determined the two mass squared
differences and the two angles, and have provided an upper bound
on the third mixing angle $\theta_{13}$.
The three generation analysis of the current data suggests 
\begin{table}
\begin{center}
\begin{tabular}{c|c|c}\hline\hline
& best fit & $3\sigma$ range   \\\hline
$\Delta m_{21}^2$ [$10^{-5}{\rm eV}^2$]  & 7.6 & 7.1 - 8.3 \\
$|\Delta m_{31}^2|$ [$10^{-3}{\rm eV}^2$]  & 2.4 & 2.0 - 2.8 \\\hline
$\sin^2\theta_{12}$ & 0.32 & 0.26 - 0.40 \\
$\sin^2\theta_{23}$ & 0.50 & 0.34 - 0.67 \\
$\sin^2\theta_{13}$ & 0.007 & $\leq$ 0.05 \\\hline\hline
\end{tabular}
\caption{The present best-fit values and the 3$\sigma$ ranges 
of oscillation parameters from \cite{Maltoni:2004ei}.
} 
\end{center}
\label{5data}
\end{table}
the $3\sigma$ ranges of the five oscillation parameters as presented
in Table 1 \cite{Maltoni:2004ei}.
Note that at present there is no constraint on any of the CP phases.

We note that, while for $\theta_{13}=0$ and $\theta_{23}=\pi/4$, 
$\mu$-$\tau$ symmetric nature is observed in 
the lepton mixing matrix, the charged-fermion masses do not
respect this symmetry at all.
If we regard the observed symmetric nature as a remnant of
some exact symmetry at some high-energy scale,
the symmetry should be broken strongly in the charged-fermion
sector, whereas it must be  broken weakly (or preserved) in the neutrino 
sector.
A nontrivial task is, therefore, to realize such asymmetric breaking
naturally \cite{chargedlepton}. 
We put aside this issue in the present work and just assume that
probable equalities among matrix elements hold only in the
neutrino sector, taking the charged-lepton mass matrix as diagonal. 
However, for the right-handed Majorana mass matrix we assume a most general 
form and do not consider it to be diagonal a priori. 

It should be emphasized that, in the following analysis, 
the flavor basis of the left-handed neutrino (the $SU(2)$ lepton doublet)
is always fixed, in such a way, that the permutations of the columns of the
Dirac mass matrix $m_D$ change physical consequences.
In general, there are 6 textures of $m_D$ which are associated with
each other by permutation of the columns.
We must take into account these 6 patterns as general possibilities,
regarding them as independent textures which lead to different predictions.

\section{The equalities among matrix elements}
\label{equal}
Before going to the study of coexistence of the equalities
and vanishing elements in the mass matrices, 
we outline the handling of the equalities among mass matrix elements 
and discuss the situation where only equalities are imposed on the neutrino 
mass matrices.

In both Dirac and Majorana mass matrices, the matrix elements are
in general complex valued.
We impose the equalities among the matrix elements  
such that these are applicable not only to the absolute
values of the matrix elements but also to the complex phases.
We will comment on the un-removable phases of the matrix elements
and CP violation of certain textures in Section 4, where we study the 
hybrid textures.

We start with the classification of the general possibilities
of the Dirac mass matrix.
It should be noted that, here and in what follows,
the textures of $m_D$ are specified by the positions of the 
matrix elements which are connected with the other elements
by equalities.
At the stage of enumerating general possibilities,
the locations of the vanishing elements specify the ``identity'' of the
textures. 
For example, the equation $(m_D)_{11}= (m_D)_{12}$ symbolizes the texture
\begin{eqnarray}
m_D = 
\begin{pmatrix}
a & a & b \\
c & d & e \\
\end{pmatrix}.
\end{eqnarray}
Since there are 6 matrix elements in the Dirac mass matrix $m_D$,
we can impose equality relations between matrix elements up to 5.
We show the complete list of the possible equalities of $m_D$
in Appendix \ref{eqDirac}.

Next let us discuss the Majorana mass matrix $M_R$.
In this work, we take $M_R$ as a $2\times 2$ 
complex symmetric matrix, which means that there are 3 independent matrix 
elements in $M_R$. 
Therefore $M_R$ can accommodate at most 2 equalities.
However, 2 equalities in $M_R$ 
imply a vanishing determinant. 
With such an $M_R$, there appears a state
which does not receive seesaw suppression in mass.
In this work, we do not consider such  spectrum and in what follows we 
will simply exclude the cases where $M_R$ has two equalities.
The three alternatives for $M_R$ (or $M_R^{-1}$) with 1 equality are: 
\begin{eqnarray}
M_R^{-1} = \begin{pmatrix}
A & B \\
B & A \\
\end{pmatrix},\quad
\begin{pmatrix}
A & A \\
A & B \\
\end{pmatrix},\quad
\begin{pmatrix}
A & B \\
B & B \\
\end{pmatrix}.
\label{1eqMR}
\end{eqnarray}
Note that in $2\times 2$ case, the equalities in $M_R$ are directly
connected to the equalities in $M_R^{-1}$.
We will examine these three textures as general possibilities
in the following discussions.
We note that for the three textures in (\ref{1eqMR}),
all the matrix elements cannot be made real by
re-definition of the right-handed neutrino fields.
Thus, although 1 equality relation reduces the number of the
free parameters by one, it does not reduce the number of the phases
which can be rotated away from the Lagrangian.
This is  different from the  case of texture zeros.
If we impose 1 zero texture in $M_R$, there is no un-removable
phase in the matrix.

Now we are in the stage to study the combination of $m_D$ and $M_R$ 
according to the total number of equalities to be distributed in them.
First of all, it is easy to see 
that the case of total 7 or 6 equalities cannot be viable
because they would either need 5 equalities 
in $m_D$ or 2 equalities in $M_R$ or both together. 

The next possibility is total 5 equalities.
In this case, there  is only one option that satisfy our 
selection criteria namely, 
\begin{itemize}

\item 4 equalities in $m_D$ and 1 equality in $M_R$

\end{itemize}
For this case, from the 7 representatives of $m_D$ in 
Appendix \ref{4eqmD} and 
3 patterns of
(\ref{1eqMR}), we have
\begin{eqnarray}
m_D &=& \begin{pmatrix}
a & a & a \\ 
a& a & b \\
\end{pmatrix},
\begin{pmatrix}
a & a & a \\ 
a& b & b \\
\end{pmatrix},
\begin{pmatrix}
a & a & b \\ 
a&  b& a \\
\end{pmatrix},
\begin{pmatrix}
a & a & b \\ 
a&  b& b \\
\end{pmatrix},
\begin{pmatrix}
a & a & b \\ 
b&  b& a \\
\end{pmatrix},\nonumber\\
M_R^{-1} &=& \begin{pmatrix}
A & B \\ B & A \\
\end{pmatrix},
\begin{pmatrix}
A & A \\ A & B \\
\end{pmatrix},
\begin{pmatrix}
A & B \\ B & B \\
\end{pmatrix}.
\label{5a}
\end{eqnarray}
Note that in (\ref{5a}), we have dropped the two Dirac mass matrices 
containing rows that are not independent of each other.
With these forms of $m_D$, 
we obtain $\mathcal{M}$ which has only one massive state
because we can rotate the right-handed fields in such a way that only one 
right-handed neutrino is coupled with the left-handed neutrinos.
We can therefore exclude these two cases from the viable possibilities. 

Note also that in (\ref{5a}), the Dirac mass matrices presented are the 
"representatives" from which all possible forms of $m_D$ are generated.
Thus (\ref{5a}) actually contains large number of the combinations.
For instance, corresponding to the first $m_D$ in (\ref{5a}), there are
5 other associated forms.
Accordingly, we must understand that there are $6 \times 3$ combinations of 
$m_D$ and $M_R$ for the first $m_D$ presented in (\ref{5a}).
However, not all of these combinations are independent.
They contain the combinations which are associated with each other
by the permutation of the two right-handed neutrinos.
Thus, it is sufficient to take account of the column exchanges
of each $m_D$ in (\ref{5a}).

Finally, we comment on the case of total 4 equalities. 
There are two possible options to be considered, for distributing 4 equalities 
in $m_D$ and $M_R$.
\begin{itemize}

\item 4 equalities in $m_D$ and 0 equality in $M_R$

\item 3 equalities in $m_D$ and 1 equality in $M_R$

\end{itemize}
These two cases cannot be excluded a priori and 
we regard them as general possibilities for total 4 equalities:
\begin{eqnarray}
m_D &=& \begin{pmatrix}
a & a & a \\ 
a& a & b \\
\end{pmatrix},
\begin{pmatrix}
a & a & a \\ 
a& b & b \\
\end{pmatrix},
\begin{pmatrix}
a & a & b \\ 
a& a & b \\
\end{pmatrix},
\begin{pmatrix}
a & a & b \\ 
a&  b& a \\
\end{pmatrix},\nonumber\\
&&\begin{pmatrix}
a & a & a \\ 
b&  b& b \\
\end{pmatrix},
\begin{pmatrix}
a & a & b \\ 
a&  b& b \\
\end{pmatrix},
\begin{pmatrix}
a & a & b \\ 
b&  b& a \\
\end{pmatrix},\nonumber\\
M_R^{-1} &=& \begin{pmatrix}
A & B \\ B & C \\
\end{pmatrix},
\label{4a}
\end{eqnarray}
and
\begin{eqnarray}
m_D &=& \begin{pmatrix}
a & a & a \\ 
b& b & c \\
\end{pmatrix},
\begin{pmatrix}
a & a & b \\ 
a& b & c \\
\end{pmatrix},
\begin{pmatrix}
a & a & b \\ 
a& c & b \\
\end{pmatrix},
\begin{pmatrix}
a & a & c \\ 
a&  b& b \\
\end{pmatrix},
\begin{pmatrix}
a & a & b \\ 
b&  c& a \\
\end{pmatrix},
\begin{pmatrix}
a & a & c \\ 
b&  b& a \\
\end{pmatrix},\nonumber\\
&&\begin{pmatrix}
a & a & a \\ 
a&  b& c \\
\end{pmatrix},
\begin{pmatrix}
a & a & b \\ 
a&  a& c \\
\end{pmatrix},
\begin{pmatrix}
a & a & b \\ 
a&  c& a \\
\end{pmatrix},
\begin{pmatrix}
a & b & c \\ 
a&  b& c \\
\end{pmatrix},
\begin{pmatrix}
a & a & c \\ 
b&  b& c \\
\end{pmatrix},
\begin{pmatrix}
a & b & c \\ 
b&  a& c \\
\end{pmatrix},\nonumber\\
&&\begin{pmatrix}
a & a & c \\ 
c&  b& b \\
\end{pmatrix},
\begin{pmatrix}
a & b & c \\ 
c&  a& b \\
\end{pmatrix},\nonumber\\
M_R^{-1} &=& \begin{pmatrix}
A & B \\ B & A \\
\end{pmatrix},
\begin{pmatrix}
A & A \\ A & B \\
\end{pmatrix},
\begin{pmatrix}
A & B \\ B & B \\
\end{pmatrix}.
\label{4b}
\end{eqnarray}
We note that there are at most 2 un-removable phases in the above
textures (\ref{4a}) and (\ref{4b}).
However, with a vanishing matrix element, the number of the un-removable phases
is reduced to one.

\section{Hybrid texture analysis}
In this section, we show the results of the combined analysis of 
the equalities and the texture zeros, according to the total number of
the reductions of the free parameters. 
One of our aim is to make a list of the realistic forms of $m_D$ and $M_R$
which have as small number of independent parameters as possible.
In other words, we search for the textures which have the strongest
predictive power with the coexistence of the vanishing elements
and the equalities among matrix elements.

Before going to the discussions, we should define the procedure 
to impose the equalities and texture zeros on the mass matrices
that we have followed.
We impose texture zeros on the mass matrices
after introducing the equalities among the matrix elements.
For instance if we consider the $m_D$ of (\ref{33ex}) and put 
$b=0$ then we get 
\begin{eqnarray}
\begin{pmatrix}
a &a&b \\
b&b&a \\
\end{pmatrix}
\,\to\,
\begin{pmatrix}
a &a&0 \\
0&0&a \\
\end{pmatrix},
\label{eqvszero1}
\end{eqnarray} 
the resultant texture belong to 4 equality and 1 zero. 
Note that we do not impose texture zeros on each entry,
but rather force the parameter $b$ to be zero.
On the other hand, if we consider (\ref{321ex}) and put $b=0$ and $c=0$
we get,  
\begin{eqnarray}
\begin{pmatrix}
a &a&c \\
b&b&a \\
\end{pmatrix}
\,\to\,
\begin{pmatrix}
a &a&0 \\
0&0&a \\
\end{pmatrix}.
\label{eqvszero2}
\end{eqnarray}
Thus, after putting the zeros, the resultant matrix is the same in both cases 
though (\ref{eqvszero2}) is obtained by setting two different parameters 
to be zero. 
In that sense (\ref{eqvszero2}) belongs to 3 equalities and 2 zeros
denoting the fact that the zeros have originated from different parameters. 
Such a classification is justified because strictly speaking,
when we impose texture zeros then it  does not imply exact
zero element but  some matrix element which is anomalously small
compared to the other elements\footnote{
From the viewpoint of 
model building it is difficult to obtain exact zeros for instance due to 
quantum corrections.}.  
Therefore in the most general scenario the two matrices can 
belong to different categories although the total number of reductions remain 
the same. 
However, it is to be noted that in our present work we have treated 
a zero as an exact zero and from this viewpoint 
both (\ref{eqvszero1}) and (\ref{eqvszero2})
will give identical results for the predictions of 
masses and mixing angles. Therefore once we consider the case of 
4 equalities and 1 zero we need not redo the calculations for 
3 equalities and 2 zeros.  
Generalizing the above we can say that in our calculations 
when we put more than one zero in any of the matrices, it eventually
increases the number of equalities of that matrix 
and reduce the number of zeros. Thus $m$ equalities and $n$ zeros   
already gets considered under $m+1$ equalities and $n-1$ zeros.
One can continue this reduction till $n=2$.  
Thus maximum number of zeros in any reduction is 2, distributed as 
one zero in $m_D$ and one zero in
$M_R$.

The maximum possible number of reductions that one can get in 
minimal seesaw model is 7.
The parameter reductions can be distributed as equalities and zeros 
according to the following tables (for total 7 and 6 reduction cases):
\begin{center}
\begin{tabular}{ccc}
Total 7 reductions & & Total 6 reductions \\
\begin{tabular}{ccc}\hline\hline
equality & zero & results \\\hline
7 & 0 & $\times$ \\\hline
6 & 1 & $\times$ \\\hline
5 & 2 & $\times$ \\\hline
4 & 3 & $\times$ \\\hline
3 & 4 & $\times$ \\\hline
$\vdots$  &$\vdots$  & $\times$ \\\hline\hline
\end{tabular} 
& &
\begin{tabular}{ccc}\hline\hline
equality & zero & results \\\hline
6 & 0 & $\times$ \\\hline
5 & 1 & $\bigcirc$ \\\hline
4 & 2 & $\bigcirc$ \\\hline
3 & 3 & $\times$ \\\hline
2 & 4 & $\times$ \\\hline
$\vdots$  &$\vdots$  & $\times$ \\\hline\hline
\end{tabular} \\
\end{tabular}
\end{center}
In the tables, the symbol ``$\bigcirc$'' means there are textures
which are compatible to the current oscillation data, and the symbol
``$\times$'' means there is no such viable one in each case.
The general textures in each case are created, for example, 
by imposing the zero elements on (\ref{5a}), (\ref{4a}) and (\ref{4b}).
By thorough examinations of all possibilities, we find three viable 
textures in 5 equalities + 1 zero and 4 equalities + 2 zero cases,
and three almost viable textures in 4 equalities + 2 zero case.
On the other hand, no viable solution exists at the level of 7 reductions.

In the 7 reductions, all possible textures amount to $\mathcal{M}$
with integer entries up to overall factor made out of the scales
in $m_D$ and $M_R$.
An interesting feature of this type of matrices is that they can
provide hierarchy among their eigenvalues by cancellation of the
numerical factors.
Since it is unlikely for the usual groups that the Clebsch-Gordan
coefficients present strong hierarchies among themselves,
the realization of the mass hierarchy along the above line gives
an insight into  model building for the charged fermion sectors
with flavor symmetry \cite{su11}.

We note that, except for some particular cases,
there are CP violating phases in each texture at
the level of 6 reductions.
Although these phases can affect physics, we have neglected
them in the texture analysis, regarding all the parameters 
in the mass matrices as real valued. 
Namely, we pick up those mass textures which can be
compatible with the data without the help of the phases.
This simplification may exclude the possibility of the textures 
which can be made viable only with nontrivial values of the phases.
A complete survey of this kind of textures needs more laborious  
calculations which is beyond the scope of this paper.

Table \ref{RES} shows the three viable and the three
``quasi viable'' textures.
Besides the 6 patterns in the table, there exist other 6 textures
which can be obtained by permuting 2-3 column of $m_D$.
Although such 6 counterparts are independent solutions, we present
only 6 textures because those predictions
are almost the same as the solutions in Table \ref{RES}.
Note also that we can create other viable solutions by relaxing 
the equalities of the solutions presented in the table.
Such daughter textures have less predictions than the original
solutions according to the number of the equalities that are relaxed. 
However, such solutions may also be of interest 
because it is practically easier to realize moderate textures
rather than the solutions in Table \ref{RES} themselves
in the context of usual model building.

In the following, we discuss the viable textures,
focusing on salient features of these solutions 
and the implications for  future experiments and leptogenesis. 

\begin{table}[t]
\begin{center}
\begin{tabular}{c|c|c|c|c|c|c|c}\hline\hline
&$m_D$, ~~~~~$M_R^{-1}$ & NH & IH & $\sin^2\theta_{12}$ &$\sin^2\theta_{13}$ & 
$\sin^2\theta_{23}$ & $|m_{ee}|$ \\\hline
I&
$
\begin{pmatrix}
a & a & a \\
0 & a & a \\
\end{pmatrix}$,
$
\begin{pmatrix}
A & A \\
A & B \\
\end{pmatrix}
$
& $\times$ & $\bigcirc$ & $\simeq \frac{1}{{3}}$ & 0 
& $\frac{1}{{2}}$ & $ \simeq \frac{1}{3}\sqrt{\Delta m^2_{31}}$ \\\hline
II&
$
\begin{pmatrix}
b & a & a \\
a & b & b \\
\end{pmatrix}$,
$
\begin{pmatrix}
A & A \\
A & 0 \\
\end{pmatrix}
$
& $\times$ & $\bigcirc$ & $\simeq {\frac{3 - \sqrt{2}}{6}}$ & 0 
& $\frac{1}{{2}}$  & $ \simeq \frac{\sqrt{2}}{3}{\Delta m^2_{31}}$ 
\\\hline
III&
$
\begin{pmatrix}
b & a & a \\
a & b & 0 \\
\end{pmatrix}$,
$
\begin{pmatrix}
0 & B \\
B & B \\
\end{pmatrix}
$
& $\times$ & $\bigcirc$ & $\simeq 0.31$ & $\simeq 0.04$ & $\simeq 0.52$
  & $ \simeq 0.527\sqrt{\Delta m^2_{31}}$  \\\hline
IV&
$
\begin{pmatrix}
a & b & a \\
0 & a & b \\
\end{pmatrix}$,
$
\begin{pmatrix}
A & 0 \\
0 & A \\
\end{pmatrix}
$
& $\bigcirc$ & $\times$ & 0.23 & 0.05 
& 0.49 & 0.0043 eV\\\hline
V&
$
\begin{pmatrix}
a & a & b \\
0 & b & a \\
\end{pmatrix}$,
$
\begin{pmatrix}
0 & B \\
B & 0 \\
\end{pmatrix}
$
& $\bigcirc$ &$\times$  & 0.24 & 0.04 & 0.49 & 0 
\\\hline
VI&
$
\begin{pmatrix}
0 & c & b \\
c & b & 0 \\
\end{pmatrix}$,
$
\begin{pmatrix}
A & A \\
A & 0 \\
\end{pmatrix}
$
& $\bigcirc$ &$\times$  & 0.19 & 0.04 & 0.50& 0  \\\hline\hline
\end{tabular}
\end{center}
\caption{The three minimal solutions (I, II, III) and the three 
``quasi viable'' solution (IV, V, VI).
The column ``NH'' and ``IH'' means the normal and the inverted 
hierarchy respectively.
The symbol ``$\bigcirc$'' means each texture can accommodate each
mass ordering, and ``$\times$'' means it cannot.
For I, II and III, the three mixing angles and the averaged mass parameter
for the neutrino-less double beta decay 
are given to the $0^{th}$ order approximation in powers of 
$\alpha = \Delta m^2_{21}/|\Delta m^2_{31}|$.
The decimals in the solutions III represent irrational factors which
have too long expressions to be shown, whereas the decimals in 
the solution IV, V and VI are the predictions with the current
best fit value of $\alpha$.
The detailed discussions for I and II are given below  (\ref{solI}) and
(\ref{solII}) and for III is found below (\ref{solIII}).}
\label{RES}
\end{table}

\subsection{The solution I}
Let us first discuss the solution I;
\begin{eqnarray}
m_D = \begin{pmatrix}
a & a & a \\
0 & a & a \\
\end{pmatrix},\quad
M_R^{-1} = \begin{pmatrix}
A & A \\
A & B \\
\end{pmatrix}.
\label{solI}
\end{eqnarray}
After the seesaw mechanism, the Majorana mass matrix
for the left-handed neutrinos becomes
\begin{eqnarray}
\mathcal{M} \,=\,
\begin{pmatrix}
A' & 2A' & 2A' \\
2A' & 3A' + B' & 3A' + B' \\
2A' & 3A' + B' & 3A' + B' \\
\end{pmatrix},
\label{solIM}
\end{eqnarray}
where we introduce the parameter $A' \equiv a^2A$ and $B' \equiv a^2B$ 
to simplify the notation. Note that this matrix obeys the so called 
scaling property between the second and third rows and second and third columns 
\cite{werner-ravi}. 
This matrix has $\mu$-$\tau$
symmetry and a zero eigenvalue such that the mass ordering is predicted 
to be inverted. 
Moreover, the reactor and the atmospheric angles are given as 
$\theta_{13} = 0$ and $\theta_{23} = 45^\circ$. 
A nontrivial consequence of this solution is thus the relation between
masses and the solar angle.
The two nonzero eigenvalues are
\begin{eqnarray}
\lambda_{\pm} = \frac{1}{2}
\left( 7A' + 2B' \pm \sqrt{ 57A'^2 + 20A'B' + 4B'^2} \right).
\end{eqnarray}
We find that these eigenvalues should be identified
as $\lambda_+ = \sqrt{|\Delta m^2_{31}| + \Delta m^2_{21} }$
and $\lambda_- = -\sqrt{|\Delta m^2_{31}|}$ in order to fit the observations.
The parameters $A'$ and $B'$ are therefore fixed in terms of the two
mass differences as
\begin{eqnarray}
A' &\simeq& - \sqrt{|\Delta m^2_{31}|}\left(
\frac{1}{3} + \frac{1}{18}\alpha + \mathcal{O}(\alpha^2)
\right),
\nonumber\\
B' &\simeq&  \sqrt{|\Delta m^2_{31}|}\left(
\frac{7}{6} + \frac{4}{9}\alpha  + \mathcal{O}(\alpha^2)
\right),
\end{eqnarray}
where $\alpha \equiv \Delta m^2_{21}/|\Delta m^2_{31}|$.
Here we are taking a combination of the solutions 
which is viable with the solar neutrino
observations.  
The solar angle is also fixed by the two mass differences.
It is given by
\begin{eqnarray}
\sin\theta_{12} \,=\,
\frac{1 + x}{\sqrt{8x^2 + (1+x)^2}},
\label{sin12}
\end{eqnarray}
where 
\begin{eqnarray}
x \equiv \frac{1}{18}\left(\sqrt{1 + \alpha} -1 - 
\sqrt{2 + 34\sqrt{1 + \alpha} + \alpha}\right),
\end{eqnarray}
By expanding (\ref{sin12}) in powers of $\alpha$, we find
\begin{eqnarray}
\sin\theta_{12} \,\simeq\, \frac{1}{\sqrt{3}} 
\,-\, \frac{1}{6\sqrt{3}}\alpha 
\,+\, \mathcal{O}(\alpha^2).
\end{eqnarray}
It is interesting to observe that, in the 0-th order, 
the solar mixing angle is predicted to be $1/\sqrt{3}$, which is the same
as in the tri-bimaximal mixing scenario. 
In fact, the whole mixing matrix $V$ can be written as
\begin{eqnarray}
V \,\simeq\, \begin{pmatrix}
-\frac{2}{\sqrt{6}} & \frac{1}{\sqrt{3}} & 0 \\
\frac{1}{\sqrt{6}} & \frac{1}{\sqrt{3}} & \frac{-1}{\sqrt{2}} \\
\frac{1}{\sqrt{6}} & \frac{1}{\sqrt{3}} & \frac{1}{\sqrt{2}} \\
\end{pmatrix}
\begin{pmatrix}
1 & \frac{\alpha}{6\sqrt{2}} & 0  \\
-\frac{\alpha}{6\sqrt{2}} & 1 & 0 \\
0 & 0 & 1 \\
\end{pmatrix}
\end{eqnarray}
up to $\mathcal{O}(\alpha)$.
Thus, this is given by the tri-bimaximal mixing matrix
multiplied by the small correction matrix.
This type of lepton mixing is naturally realized by a class
of flavor models which utilize the Scherk-Schwarz twisted boundary
conditions \cite{SS}.
It might be interesting to study this type of deviation from the 
tri-bimaximal mixing and its implication for  future neutrino experiments 
systematically.

The effective neutrino mass $|m_{ee}|$ which is responsible for 
the neutrino-less double beta decay is given by the element $A'$
itself.
Thus in this texture we find $|m_{ee}|$ is predicted as
\begin{eqnarray}
|m_{ee}| \, \simeq \, \frac{1}{3}\sqrt{|\Delta m^2_{31}|} \,=\, 0.016
\quad {\rm eV}
\end{eqnarray}
at the best fit of the mass difference.
With this value, 
neutrino-less double beta decay may be detectable 
in forthcoming experiments.

Since there are only two mixing angles in $V$, CP must
be conserved at low energy.
However, at high energy there is one unremovable phase 
which can violate CP. 
Such CP violation is conveniently measured by
the weak basis invariant 
$I_h = {\rm Im}{\rm Tr}\Big[ hHM_R^* h^* M_R \Big]$
where $h \equiv m_D^* m_D^{\rm T}$ and $H \equiv M_R^\dag M_R$~\cite{CPhi}.
For the texture I, we find
\begin{eqnarray}
I_h = \frac{a^4 ( A^2 - 4|A-B|^2 )B\sin\phi_B}{|A-B|^4 A^3}
\end{eqnarray}
in the basis where only $B$ is complex; $B \to B e^{i\phi_B} $.
The leptogenesis \cite{leptogenesis1} is thus possible, 
with the lepton asymmetry proportional to this quantity.

\subsection{The solution II}
Next we discuss the solution II;
\begin{eqnarray}
m_D = \begin{pmatrix}
b & a & a \\
a & b & b \\
\end{pmatrix},\quad
M_R^{-1} = \begin{pmatrix}
A & A \\
A & 0 \\
\end{pmatrix}.
\label{solII}
\end{eqnarray}
After the seesaw mechanism, the Majorana mass matrix
for the left-handed neutrinos becomes
\begin{eqnarray}
\mathcal{M}\, = \,
\begin{pmatrix}
b'(2a' + b') & a'^2 + a'b' + b'^2 & a'^2 + a'b' + b'^2 \\
a'^2 + a'b' + b'^2 & a'(a'+2b') & a'(a'+2b') \\
a'^2 + a'b' + b'^2 & a'(a'+2b') & a'(a'+2b') \\
\end{pmatrix},
\label{solIIM}
\end{eqnarray}
where we define $a' \equiv a\sqrt{A}$ and $b' \equiv b\sqrt{A}$
to simplify  the notation.
As in the previous case, this matrix also has $\mu$-$\tau$
symmetry and a zero eigenvalues.
The mass hierarchy is thus predicted as inverted ordering, 
and the reactor and the atmospheric angles are given as 
$\theta_{13} = 0$ and $\theta_{23} = 45^\circ$.
The two nonzero eigenvalues are
\begin{eqnarray}
\lambda_{\pm} = \frac{1}{2}
\left( 2a'^2 + 6a'b' + b'^2 \pm \sqrt{3}\sqrt{ 
4a'^4 + 8a'^3b' + 8a'^2b'^2 + 4a'b'^3 + 3b'^4} \right),
\end{eqnarray}
We find that these eigenvalues should be identified as 
$\lambda_+ = \sqrt{|\Delta m^2_{31}| + \Delta m^2_{21} }$
and $\lambda_- = -\sqrt{|\Delta m^2_{31}|}$ in order to fit the observations.
A solution of this equation system is found to be
\begin{eqnarray}
&&a' \,\simeq\, \left( |\Delta m^2_{31}|\right)^{1/4}\left(
 \frac{(3 + \sqrt{7})\sqrt{-2 + \sqrt{7} }}{3 \cdot 2^{3/4}}
\,+\, \frac{7\sqrt{14} + 16\sqrt{7} + 7\sqrt{2} - 56}{336 \cdot 2^{1/4} 
\sqrt{-2 + \sqrt{7}}}\, \alpha \,+\, \mathcal{O}(\alpha^2) \right),\nonumber\\
&&b' \,\simeq\, \left( |\Delta m^2_{31}|\right)^{1/4}\left(
-\frac{2^{1/4}\sqrt{-2 + \sqrt{7}}}{3}
\,+\, \frac{-7\sqrt{14} + 2\sqrt{7} + 14\sqrt{2} +14}{168 \cdot 2^{1/4} 
\sqrt{-2 + \sqrt{7}}}\, \alpha \,+\, \mathcal{O}(\alpha^2)\right).\nonumber\\
\label{abII}
\end{eqnarray}
Here we show approximate expressions for $a'$ and $b'$ because
the exact expressions are too long and complicated to present here.
The solar angle is also fixed by the two mass differences.
It is written in terms of $a'$ and $b'$, and the mass differences as
\begin{eqnarray}
\sin\theta_{12} &\,=\,&
\left| \frac{ -2a'(a' + 2b') + \lambda_+ }{\sqrt{
2(a'^2 + a'b' + b'^2 )^2 + (-2a'(a' + 2b') + \lambda_+)^2 }}
\right|\nonumber\\
&\simeq & \sqrt{\frac{3 - \sqrt{2}}{6}}
\,-\, \frac{3\sqrt{2} - 2}{16\sqrt{3}(3 - \sqrt{2})^{3/2}}\alpha 
\,+\, \mathcal{O}(\alpha^2)\nonumber\\
&\,=\,& 0.514 - 0.0405 \, \alpha + 0.0258 \, \alpha^2 \,+\, 
\mathcal{O}(\alpha^3).
\label{sin12b}
\end{eqnarray}
We can see that the prediction for the solar angle agrees with
the current $3\sigma$ range although it is close to its lower bound.

The averaged neutrino mass $|m_{ee}|$ which is responsible for 
the neutrino-less double beta decay is  given by $b'(2a' + b')$
in this case.
We find $|m_{ee}|$ is predicted as
\begin{eqnarray}
|m_{ee}| \, \simeq \, \frac{\sqrt{2}}{3}\sqrt{|\Delta m^2_{31}|} \,=\, 0.023
\quad {\rm eV}
\end{eqnarray}
at the best fit of the mass difference and a signal can be expected in 
future 
We have a slightly better chance to detect
neutrino-less double beta decay 
measurements. 

CP violation is possible at high-energy as the weak-basis invariant
$I_h$ is non-vanishing. 
In fact, it becomes
\begin{eqnarray}
I_h = \frac{2ab\left( 
a^2 - b^2 -3ab\cos\phi_b \right)\sin\phi_b}{A^4}
\end{eqnarray}
in the basis where only $b$ is complex; $b \to b e^{i\phi_b}$.
On the other hand, no CP violation occurs in the neutrino oscillation
because of the vanishing $\theta_{13}$.

\subsection{The solution III}
Finally let us discuss the solution III.
The viable texture is given by
\begin{eqnarray}
m_D = \begin{pmatrix}
b & a & a \\
a & b & 0 \\
\end{pmatrix},\quad
M_R^{-1} = \begin{pmatrix}
0 & B \\
B & B \\
\end{pmatrix}.
\label{solIII}
\end{eqnarray}
After the seesaw mechanism, the Majorana mass matrix
for the left-handed neutrinos becomes
\begin{eqnarray}
\mathcal{M} \,=\,
\begin{pmatrix}
a'(a'+2b') & a'^2 + a'b' + b'^2 & a'^2 \\
a'^2 + a'b' + b'^2 & b'(2a' + b') & a'b' \\
a'^2 & a'b' & 0 \\
\end{pmatrix},
\label{MIII}
\end{eqnarray}
where we introduce the parameter $a' \equiv a\sqrt{B}$ and 
$b' \equiv b\sqrt{B}$ to simplify the notation.
This has one zero element in the final matrix and has been discussed 
in \cite{werner-merle}.
Unlike the two solutions found in the previous subsection,
there is no  $\mu$-$\tau$ symmetry in (\ref{MIII}).
Therefore the reactor and the atmospheric angles no longer satisfy
$\theta_{13} = 0$ and $\theta_{23} = 45^\circ$, and all nonzero values 
of the mixing angles can be described as functions of the mass differences.
We note that, because of the lack of the $\mu$-$\tau$ symmetry, there appears
an additional solution in this case.
Namely, the $m_D$ which is obtained by 2-3 column exchange of 
(\ref{solIII}) can also be compatible with the data.
Although these two solutions are independent in the sense that they
are not related to each other by basis rotations, their physical consequences
are almost the same.
We thus discuss only (\ref{solIII}) to illustrate
important features of these solutions.

The two nonzero eigenvalues of (\ref{MIII}) are
\begin{eqnarray}
\lambda_{\pm} = \frac{1}{2}
\left( a'^2 + 4a'b' + b'^2 \pm \sqrt{ 9a'^4 + 8a'^3b' + 14a'^2b'^2 + 8a'b'^3
 + 5b'^4} \right).
\end{eqnarray}
We find that these eigenvalues should be identified
as $\lambda_- = -\sqrt{|\Delta m^2_{31}| + \Delta m^2_{21} }$
and $\lambda_+ = \sqrt{|\Delta m^2_{31}|}$
in order to fit the observations.
The parameters $a'$ and $b'$ are therefore fixed in terms of the two
mass differences.
However, the exact expressions are too long and complicated and 
it is not appropriate to present them here.
Instead, we shall approximate them in powers of 
$\alpha$;
\begin{eqnarray}
a' &\simeq & - \left(|\Delta m^2_{31}\right|)^{1/4}\left(
0.848 + 0.116\,\alpha  + \mathcal{O}(\alpha^2)
\right),
\nonumber\\
b' &\simeq&  \left(|\Delta m^2_{31}\right|)^{1/4}\left(
0.227 + 0.201\,\alpha + \mathcal{O}(\alpha^2)
\right).
\end{eqnarray}
Here we are picking up a combination of the solutions for 
$\lambda_- = -\sqrt{|\Delta m^2_{31}| + \Delta m^2_{21} }$
and $\lambda_+ = \sqrt{|\Delta m^2_{31}|}$, which gives the correct 
mixing angles.
Note that we represent the leading and the first order coefficients
by decimal numbers.
These terms can also be represented as functions of integers
as in (\ref{abII}), but the expressions are not so simple as
in the case of the other two solutions.
The three angles are also fixed by the two mass differences.
They are given by
\begin{eqnarray}
&&\sin\theta_{12} \,\simeq\, 0.561 + 0.00775\,\alpha + \mathcal{O}(\alpha^2),\\
&&\sin\theta_{23} \,\simeq\, 0.719 + 0.0171\,\alpha + \mathcal{O}(\alpha^2),\\
&&\sin\theta_{13} \,\simeq\, 0.193 + 0.149\,\alpha + \mathcal{O}(\alpha^2).
\end{eqnarray}
Here we note again that the decimals in the above 
expressions represent definite irrational numbers
which are too long and complicated to be presented in closed form.
It is interesting that the predictions for $\theta_{12}$ and $\theta_{23}$ 
are very close to the current best fit values indicated in Table 1.
We see that this texture predicts a relatively large $\theta_{13}$ which 
can be measured in forthcoming experiments like Double-Chooz. 

The effective neutrino mass $|m_{ee}|$ which is responsible for 
the neutrino-less double beta decay is given by 
$|m_{ee}| = |a'(a' + 2b')|$ and  
is predicted as
\begin{eqnarray}
|m_{ee}| \, \simeq \, 0.527\sqrt{|\Delta m^2_{31}|} \,=\, 0.026
\quad {\rm eV}
\end{eqnarray}
at the best fit of the mass difference.
Thus, neutrino-less double beta decay will be detectable 
in forthcoming experiments.

Finally we comment on CP violation.
Since $\theta_{13}$ is nonzero with the solution III, 
one might expect that there exists corelation 
between CP violation phenomena at high and low-energy scale.
However, this is not the case. 
The invariant measure $I_h$ is given by
$I_h = {\rm Im}\Big[  h_{12}^2 -2h_{11}h_{12}^*   -2h_{12}h_{22} \Big]/|B|^4$,
but $h_{12}$ is real valued for the $m_D$ in (\ref{solIII}).
Thus the leptogenesis does not occur with the solution III.
On the other hand, the phase $\delta$ in the PMNS matrix
is not vanishing.
The CP violation caused by $\delta$ can be conveniently described with,
for example, $I_l = {\rm Tr}\Big[ \mathcal{M}\mathcal{M}^\dag, m_l m_l^\dag
\Big]^3$, where $m_l$ is the charged-lepton mass matrix \cite{CPlow}.
It is easily seen that the basis-independent quantity $I_l$
takes the form $I_l = C_1 \cdot {\rm Im}\Big[ a b^* \Big] + 
C_2 \cdot {\rm Im}\Big[ a^2 b^*{}^2 \Big]$, where $C_1$ and $C_2$
are nonzero real values, and $I_l$ is therefore nonzero in general.
To summarize, with the solution III, CP violation in the oscillation
can be measured, while leptogenesis is not possible.

\subsection{The solution IV, V and VI}
Interestingly, the three solutions I, II and III are consistent only 
with the inverted mass ordering for neutrino mass spectrum.
From this fact, we conclude that the hybrid property prefers 
the inverted hierarchy.
However, there exist three quasi viable textures IV, V and VI 
with normal hierarchy.
We call them quasi-viable as 
certain predictions of these textures are marginally 
consistent with the $3\sigma$ range presented in Table 1,
This can be seen, for example, with V and VI in which 
the element $\mathcal{M}_{11}$ is vanishing.
As a general consequence of $\mathcal{M}_{11}= 0$ and the normal
hierarchy with $m_1 = 0$, the solar and reactor angles 
correlate  with $\alpha$ as $\tan\theta_{13}/\sin\theta_{12} = \alpha^{1/4}$.
This relation needs small $\theta_{12}$ and $\alpha$, and large (just below
the current $3\sigma$ bound) $\theta_{13}$.

With the solutions IV and V, there are is no CP violating phase
for the right-handed mass matrix  
since the heavy neutrino are degenerate in mass.
Such degeneracy can be relaxed by some corrections, for example,
renormalization group evolution from the texture scale (at which the textures
are assumed) down to the right-handed neutrino scale.
However, even with relaxations of the degeneracy, 
the lepton asymmetries for IV and V are small because the off-diagonal
components of $m_D^*m_D^{\rm T}$ are real valued.
It may be possible to generate a non-zero value for this by introducing 
two-loop renormalization group effects \cite{babu}. In this paper 
we do not consider these effects. 

On the other hand, we have nonzero $I_h$ and $I_l$ within the solution VI.
Therefore the texture VI can give rise to a connection 
between leptogenesis and CP violation in neutrino oscillation.
We discuss this connection in detail in the next section.

\begin{figure}
\begin{center}
\scalebox{1.1}{
\includegraphics{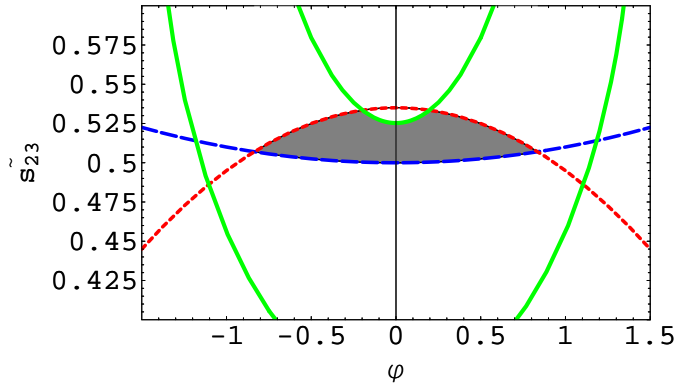}
}
\end{center}
\caption{Constraints from the low-energy data (Table 1)
on the $\varphi$-$\tilde{s}_{23}$ plane.
The green (solid) curves shows the upper and lower
bounds for the mass ratio $\alpha$.
The blue (dashed) and red (dotted) curves correspond to
the upper bound for $\theta_{13}$ and lower bound for $\theta_{12}$,
respectively.
The shaded region is allowed.}
\label{range}
\end{figure}

\section{CP violation at high and low-energy scales}
With the solution VI, both $I_h$ and $I_l$ are non-vanishing,
and there is a correlation between CP violation phenomena at
high and low-energy scales.
It is interesting if we can predict the amount of CP violation 
at low energy in terms of the phase of the Yukawa coupling responsible 
for successful baryogenesis via leptogenesis.
In this section, we address this issue and study the  connection between 
leptogenesis
and CP violation at low energy with the interesting 
example of the solution VI.

\begin{figure}
\begin{center}
\scalebox{1.2}{
\includegraphics{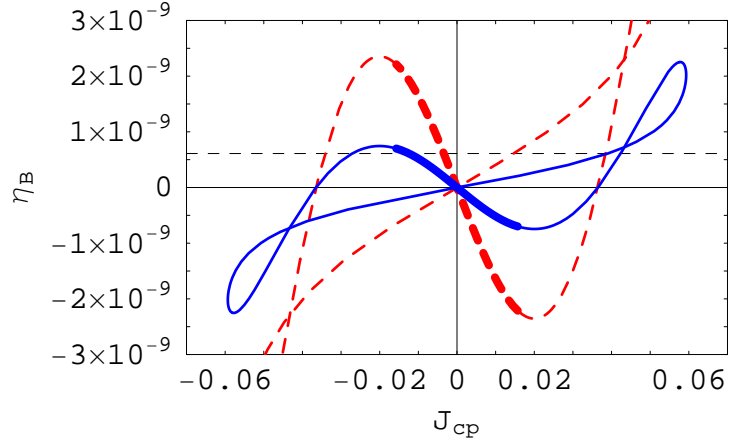}
}
\end{center}
\caption{A parametric plot for the baryon-to-photon ratio $\eta_B(\varphi)$ 
and the Jarlskog invariant $J_{\rm CP}(\varphi)$, as functions of the
phase parameter $\varphi$.
The Horizontal dashed line is the constraint by the WMAP observation 
$\eta_B = 6 \times 10^{-10}$~\cite{Spergel:2006hy}.
The dashed curve is for the heavy neutrino scale 
$A = 10^{-13.5} \, {\rm GeV^{-1}}$,
and the solid curve is for $A = 10^{-13} \, {\rm GeV^{-1}}$.
For each curve, the thick line corresponds to the parameter range
$-0.7 < \varphi < 0.7$ where the low-energy observables are consistent
with the data, whereas the thin curve corresponds to the other region.}
\label{JcpetaB}
\end{figure}
To perform this program, it is convenient to parameterize 
the Dirac mass matrix $m_D$ as \cite{sahu}
\begin{eqnarray}
m_D = \begin{pmatrix}
0 & 0 & x e^{i\varphi}  \\
0 & y & z e^{i\varphi'} \\
\end{pmatrix}
\widetilde{V}^T \cdot \widetilde{P},
\label{sahu}
\end{eqnarray}
where $x,y,z$ are positive-real parameters, and $\widetilde{V}$ is an unitary
matrix which contains only one Dirac type phase (conveniently
parameterized as (\ref{pmns}) without Majorana phases), and
$\widetilde{P}$ is a diagonal phase matrix.
Taking account of the texture zeros and equalities in $m_D$
of the solution VI, the general form (\ref{sahu}) is reduced to 
\begin{eqnarray}
m_D = Z^{\rm T}\cdot \widetilde{V}^T \cdot \widetilde{P},
\label{sahu2}
\end{eqnarray}
where $\widetilde{P} = {\rm diag}( e^{i 2\varphi},  e^{i\varphi}, 1)$ and
\begin{eqnarray}
Z^{\rm T} &\,=\,& x \begin{pmatrix}
0 & 0 &  e^{i\varphi}  \\
0 & \tilde{s}_{23}/\tilde{s}_{12} & \tilde{s}_{23}\tilde{t}_{23}/\tilde{t}_{12} \\
\end{pmatrix}, \\
\widetilde{V}
&\,=\,&
\begin{pmatrix}
1 & & \\
& \tilde{c}_{23} & \tilde{s}_{23} \\
&-\tilde{s}_{23} & \tilde{c}_{23} \\
\end{pmatrix}
\begin{pmatrix}
\tilde{c}_{12} & \tilde{s}_{12} &  \\
-\tilde{s}_{12} & \tilde{c}_{12} & \\
 & & 1 \\
\end{pmatrix},
\end{eqnarray}
where $\tilde{s}_{12} = \tilde{s}_{23}/\sqrt{\tilde{s}_{23}^2 + \tilde{c}_{23}^4}$.
Therefore (\ref{sahu2}) contains two real parameter $x$ and 
$\tilde{\theta}_{23}$, and one phase $\varphi$, as it should do.
With the parameterization (\ref{sahu2}), the seesaw formula is
written as
\begin{eqnarray}
\mathcal{M} \,=\,
-\widetilde{P} \widetilde{V} K \widetilde{V}^{\rm T} \widetilde{P},
\end{eqnarray}
where $K$ is a complex-symmetric $3\times 3$ matrix of 
$K = Z M_R^{-1} Z^{\rm T}$, but it has nonzero elements only 
in the lower-right $2 \times 2$ block.
This is an advantage of the form (\ref{sahu}) because
diagonalization of a $2\times 2$ matrix is significantly 
easier than $3 \times 3$ case.
The PMNS matrix is given by
$V^* = \widetilde{P}\widetilde{V} U$ where $U$ is a unitary matrix which 
diagonalize
$K$ such that $K = U \cdot {\rm diag}(0, m_2, m_3) \cdot U^{\rm T}$.

We should identify the two mass eigenvalues $m_{2,3}$ as
$m_2 = x^2A \lambda_2 = \sqrt{\Delta m^2_{21}}$ and 
$m_3 = x^2A \lambda_3 = \sqrt{\Delta m^2_{31}}$,
where $\lambda_{2,3}$ is the eigenvalue of $K/(x^2A)$ which is a function 
of $\tilde{s}_{23}$ and $\varphi$. 
Thus the mass ratio $\lambda_2/\lambda_3 = \sqrt{\alpha}$ gives
a constraint on the $\tilde{s}_{23}$-$\varphi$ space.
On the other hand, the overall scale of the neutrino mass
gives a relation between $x$ and $A$ as
$x = (\sqrt{\Delta m^2_{31}}/(A\lambda_3))^{1/2}$.
The three angle and the mass ratio $\alpha$ are controlled
by only two parameters $\tilde{s}_{23}$ and $\varphi$.

Since the analytic expressions for the low-energy observables
are complicated, we check the constraints on $\varphi$-$\tilde{s}_{23}$
space numerically.
Fig.~\ref{range} shows the allowed region on $\varphi$-$\tilde{s}_{23}$
plane.
The upper(lower) green(solid) curves are the $3\sigma$ lower(upper) bound
of the ratio $\alpha$, which are obtained from 
$\lambda_2/\lambda_3 = \sqrt{\alpha}$.
More important constraints come from the reactor bound 
$\sin\theta_{13} < 0.22$ and the lower bound of the solar angle
$\sin\theta_{12} > 0.51$, which are shown by red(dotted) and blue(dashed)
curves respectively.
Both $\theta_{13}$ and $\theta_{12}$ get increased as 
$\tilde{s}_{23}$ increases, so that the shaded region remains allowed
by the current oscillation data.
The upper bound on $\theta_{12}$ draws a curve above the shaded region
and it does not reduce the allowed space.
In the following, we fix $\tilde{s}_{23} = 0.52$ to assess
maximal impact on the low-energy CP violation.
For $\tilde{s}_{23} = 0.52$,  
the possible range of the phase $\varphi$ is $-0.7< \varphi < 0.7$.
With these parameters, the three mixing angles are predicted as
$\sin\theta_{23} = 0.74$, $\sin\theta_{12} = 0.53-0.54$,
$\sin\theta_{13} = 0.21-0.22$ and $\alpha = 0.025-0.030$.

The Dirac type phase $\delta$ in $V$ can be measured in long-baseline
experiments \cite{LB}.
The CP violation arises in the difference of transition probability
$P(\nu_e \to \nu_\mu)- P(\bar{\nu}_e \to \bar{\nu}_\mu)$.
The difference is proportional to the leptonic version of
the Jarlskog invariant \cite{Jcp}
\begin{eqnarray}
J_{\rm CP} = {\rm Im}[ V_{e1}V_{e2}^* V_{\mu 1}^* V_{\mu 2} ].
\end{eqnarray}
As discussed above, the mixing matrix $V$ depends only on the two parameters
$\tilde{s}_{23}$ and $\varphi$. 
With  $\tilde{s}_{23} = 0.52$, $J_{\rm CP}$ is bounded as
$-0.02 \lesssim J_{\rm CP} \lesssim 0.02$ 
for $-0.7< \varphi < 0.7$.

The CP asymmetry $\epsilon$ with the heavy neutrino decay (lighter one)
is given by
\begin{eqnarray}
\epsilon_1 \,\simeq \, \frac{1}{8\pi} 
\cdot
\frac{{\rm Im}\Big[ (\bar{m}_D^* \bar{m}_D^{\rm T})_{12}^2 \Big]}
{(\bar{m}_D^* \bar{m}_D^{\rm T})_{11}v^2}
\cdot
f \! \left(\frac{M_2^2}{M_1^2}\right),
\end{eqnarray}
where $f(x) = \sqrt{x}(1 - (1 + x)\ln[(1+x)/x])$ and   
$v = 174 \,{\rm GeV}$ is the vacuum expectation value of the Higgs field.
The right-handed neutrino masses are denoted by
$M_1$ and $M_2$ with $M_1 = (\sqrt{5} - 1)/2A$ and $M_2 = (\sqrt{5} + 1)/2A$.
We neglect the contribution from the self-energy diagram which
is small compared to the vertex one for the heavy neutrino scale
of $\gtrsim 10^9 \, {\rm GeV}$ in hierarchical case. 
In the above, the Dirac mass matrix $\bar{m}_D$ is in  
the basis where $M_R$ is diagonal.
The baryon-to-photon ratio is given by
\begin{eqnarray}
\eta_B \simeq - 10^{-2}\epsilon_1 k_{\rm f},
\end{eqnarray}
where the factor $10^{-2}$ represents the sphaleron conversion
and the dilution due to the photon productions from the
onset of leptogenesis until recombination.
The factor $k_{\rm f}$ is the final efficiency factor which we
are taking $k_{\rm f} = 2.0 \times 10^{-2}$ in this case.

Fig.~\ref{JcpetaB} shows $\eta_B$ and $J_{\rm CP}$ as a parametric
plot with respect to $\varphi$.
We put details about the plot in the caption.
We can see a sharp correlation between $\eta_B$ and $J_{\rm CP}$.
In particular, the sign of $J_{CP}$ is predicted to be negative;
the disappearance probability of the anti-neutrino $\bar{\nu}_e$ 
will be observed greater than that of the ordinary $\nu_e$.
It is also clear that there is a lower bound of the mass scale
$A^{-1}$ as $A^{-1} \gtrsim 10^{13} \, {\rm GeV}$.
If the CP violation $-0.02 \lesssim J_{CP} < 0$ is measured,
then it is an indirect measurement of the right-handed mass
scale under the assumption that the leptogenesis is solely
responsible for the baryon asymmetry of the universe.

\section{Renormalization Group Effects} 

It is to be noted that the Majorana mass matrices obtained through 
seesaw diagonalization are implicitly 
at some high scale which depends on the
mass of the heavy neutrinos.
Consequently the mixing angles and the mass eigenvalues
are the corresponding
quantities at the high scale.
To obtain the values at the low scale, renormalization group (RG)
induced running effects need to be incorporated \cite{rgpapers}.
Impact of RG running with  tri-bimaximal mixing at high scale
has been considered in \cite{plentinger,ourpapers}. It was found that
these effects are typically small for hierarchical spectrum.
Considerable running can be possible for quasi-degenerate
neutrinos depending on the values of Majorana phases \cite{rgqlc}.
However since the RG induced corrections to the mass matrix elements 
are multiplicative in nature it is expected that a zero in  the mass matrix 
$\cal{M}$ will remain a zero \cite{rgpapers}. 
It is also shown in \cite{werner-ravi} that a mass matrix obeying 
scaling properties are stable against RG corrections. 
Therefore it is plausible that the textures which we find as allowed 
will be stable against RG corrections. 
However it is possible that certain textures which are disallowed 
marginally may get allowed if one included RG effects. 
In this paper we do not attempt to classify such textures. 
Renormalization effects for texture zero mass matrices have
been discussed in \cite{mnuRGE}. In particular,
\cite{hagedorn} discussed radiative generation and stability of
texture zeros in the context of type-I seesaw models for running from 
low to high scale 
and reached the same conclusion that the RG effects cannot make an allowed
texture
forbidden but the converse may be possible. Thus we do not expect
the allowed patterns to get excluded by RG effects.

\section{Conclusions }

In this work we consider simultaneous presence of
equalities and texture zeros in the elements of Dirac and
Majorana mass matrices
in the context of the minimal seesaw model containing two heavy right-handed 
neutrinos.
It is well known that because of the symmetric nature of the Majorana 
Mass matrix ($M_R$) the off-diagonal elements are equal. 
In the present study, we impose additional equalities among the elements of the 
Majorana mass matrix as well as on the elements of the 
Dirac mass matrix ($m_D$)  at some high scale.
Equalities among matrix elements of neutrino mass matrices can arise for 
example due to $\mu$-$\tau$ exchange symmetry which predicts 
$\theta_{13}$=0 and  
$\theta_{23} = \pi/4$
in the basis where the charged lepton mass matrix is diagonal.
Such equalities reduce the number of free parameters in the
theory and hence increase its predictive power.
Another way to reduce the number of free parameters is the postulation of
texture zeros which can also be motivated by certain class of flavor
symmetries in the mass matrix.

We classify and enumerate the general possibilities of mass matrices
with equalities among  elements.
Then we perform a hybrid texture analysis combining both
equalities and zeros.
Our aim is to identify the left-handed Majorana mass matrices obtained by
seesaw diagonalization, that are compatible with the neutrino oscillation
data.
We study a large number of independent options (more than 400) and find that
at the level of minimal number of free parameters ({\it i.e.}
with maximum number of conditions imposed
on the elements of the  Dirac and Majorana mass matrices), 
only 6 textures stand out to be consistent with global 
neutrino oscillation data.
These 6 patterns, presented in Table 2 are thus, quite special 
and rare from the point of view of the parameter sets realized in nature.
These textures are characterized by two free parameters (ignoring the phases)
so that 
there exist 3 relations among 5 oscillation parameters in each solution.
We formulate these  relations by taking the two mass squared differences as
input and 3 mixing angles as output parameters.
In two out of the three cases the elements of the PMNS matrix  
are found to be given by irrational but simple algebraic numbers, 
to the 
leading order in the small parameter 
$\alpha = \Delta m^2_{21}/|\Delta m^2_{31}|$. 

All the six solutions in Table 2 have one physical phase. 
We study the  
the possibility of obtaining leptogenesis in these models and 
explore if there is any connection between the phase responsible for 
generation of lepton asymmetry and the low energy CP phase. 
We find that there is only one solution in which a connection between 
leptogenesis and 
low energy CP violation is possible ignoring radiative effects. 

It is interesting to observe that the first 3 solutions in Table 2 
are consistent with the data only with the inverted mass hierarchy.
A priori, there is no reason that some texture must belong to a 
particular hierarchy.
The basic principles which we take are the equalities among matrix elements, 
texture zeros and minimality of the parameters. 
Thus we conclude that the minimal seesaw mechanism prefers the inverted
hierarchy under the constraints which are likely to stem from physics 
beyond the standard model.
In fact, many authors have tried to explain the generation structure
invoking discrete or other symmetries, where the equalities and
vanishing elements in Yukawa couplings are often realized as 
direct consequences of imposed flavor symmetries or secondary products
of model constructions \cite{S3tex}.
While the inverted hierarchy seems somewhat special in the sense
that it shows sharp contrast to all the other fermions,
the nature seems to be open to the inverted hierarchy in the context of  
hybrid texture.

\vspace{2mm}
\subsection*{Acknowledgments}
The authors thank A. Mohanty for a careful reading of the draft.  
S.G. and A.W. acknowledges support from the Neutrino Project under the
XIth plan of Harish-Chandra Research Institute.

\appendix
\section{The equalities in the Dirac mass matrix}
\label{eqDirac}
In this appendix, we show the detailed classification 
of the Dirac mass matrix $m_D$.
Since $m_D$ has six entries, we can impose equalities on $m_D$ up to five.
\subsection{1 equality}
We shall start with 1 equality in $m_D$.
By imposing 1 equality among 6 matrix elements, the 6 elements
are divided into 5 groups, that is, for instance 
$(m_{D})_{11} = (m_{D})_{12}$, and other 4 matrix elements. 
This situation can be symbolized by $(2,1,1,1,1)$, where each entry means
the ``slot'' of the independent parameter.
Since we impose 1 equality among 6 elements, the number of the 
independent parameters is reduced to 5. 
Therefore we have 5 entries in $(2,1,1,1,1)$.
The number of each entry in the first bracket
denotes the number of the matrix elements
included in each group.
The sum of the entries must be equal to 6.

The $(2,1,1,1,1)$ case includes ${}^6 {\rm C}_2 = 15$ patterns of 
different textures.
The ``representatives'' are
\begin{eqnarray}
&&(m_D)_{11} = (m_D)_{21}
\quad \to \quad {\rm 3\,\, patterns}\\
&&(m_D)_{11} = (m_D)_{12}
\quad \to \quad {\rm 6\,\, patterns}\\
&&(m_D)_{11} = (m_D)_{22}
\quad \to \quad {\rm 6\,\, patterns}
\end{eqnarray}
Here ``representative'' means that
the other patterns can be generated by the permutation of
the rows and the columns from the above three matrices.
In other words, the above three matrices are not related to each
other by permutations of the rows and the column, so that they
compose a set of ``primary'' matrices in this category.

\subsection{2 equalities}
Here we consider 2 equalities in $m_D$.
Since we have 2 equalities, the matrix elements are divided into
4 groups.
There are two types of distributions;
(2,2,1,1) and (3,1,1,1).

\paragraph{(2,2,1,1) case} 
In this case, there are ${}^6 {\rm C}_2 \times {}^4 {\rm C}_2 = 90$
mass matrices.
If we regard the first two groups of (2,2,1,1) as identical, then
the total number is reduced to $90/2 = 45$ patterns.
The representatives are
\begin{eqnarray}
&&(m_D)_{11} = (m_D)_{21},\quad (m_D)_{12} = (m_D)_{22}
\quad \to \quad {\rm 3\,\, patterns}\\
&&(m_D)_{11} = (m_D)_{12},\quad (m_D)_{21} = (m_D)_{22}
\quad \to \quad {\rm 3\,\, patterns}\\
&&(m_D)_{11} = (m_D)_{22},\quad (m_D)_{12} = (m_D)_{21}
\quad \to \quad {\rm 3\,\, patterns}\\
&&(m_D)_{11} = (m_D)_{12},\quad (m_D)_{22} = (m_D)_{23}
\quad \to \quad {\rm 6\,\, patterns}\\
&&(m_D)_{11} = (m_D)_{22},\quad (m_D)_{12} = (m_D)_{23}
\quad \to \quad {\rm 6\,\, patterns}\\
&&(m_D)_{11} = (m_D)_{23},\quad (m_D)_{12} = (m_D)_{22}
\quad \to \quad {\rm 6\,\, patterns}\\
&&(m_D)_{11} = (m_D)_{21},\quad (m_D)_{22} = (m_D)_{23}
\quad \to \quad {\rm 6\,\, patterns}\\
&&(m_D)_{11} = (m_D)_{22},\quad (m_D)_{21} = (m_D)_{23}
\quad \to \quad {\rm 6\,\, patterns}\\
&&(m_D)_{11} = (m_D)_{23},\quad (m_D)_{21} = (m_D)_{22}
\quad \to \quad {\rm 6\,\, patterns}
\end{eqnarray}
All 45 patterns can be generated from these 9 patterns.
It should be noted again that we regard the textures which 
is related by the label exchange of the first two entries of (2,2,1,1)
as identical.
The classification of the above 9 patterns is similar to the
general possibilities for the 2 zero textures for $m_D$.

\paragraph{(3,1,1,1) case} 

We have ${}^6 {\rm C}_3 = 20$ general possibilities and
three representatives in this category.
\begin{eqnarray}
&&(m_D)_{11} = (m_D)_{12}= (m_D)_{21}
\quad \to \quad {\rm 12\,\, patterns}\\
&&(m_D)_{11} = (m_D)_{12}= (m_D)_{13}
\quad \to \quad {\rm 2\,\, patterns}\\
&&(m_D)_{11} = (m_D)_{22}= (m_D)_{23}
\quad \to \quad {\rm 6\,\, patterns}
\end{eqnarray}
All 20 patterns can be generated from these 3 patterns.
An easy way to understand these 3 patterns comes from the analogy
with the 3 zero textures in $m_D$.

\subsection{3 equalities}
Here we consider 3 equalities in $m_D$.
Since we have 3 equalities, the matrix elements are divided into
3 groups.
There are three types of distributions;
(3,2,1), (4,1,1) and (2,2,2).
Let us see in turn.

\paragraph{(3,2,1) case} 
In this case, there are ${}^6 {\rm C}_3 \times {}^3 {\rm C}_2 = 60$
patterns of textures.
The representatives are
\begin{eqnarray}
&&(m_D)_{11}=(m_D)_{12}=(m_D)_{13},\quad(m_D)_{21}=(m_D)_{22}
\quad \to \quad {\rm 6\,\, patterns}\\
&&(m_D)_{11}=(m_D)_{12}=(m_D)_{21},\quad(m_D)_{13}=(m_D)_{22}
\quad \to \quad {\rm 12\,\, patterns}\\
&&(m_D)_{11}=(m_D)_{12}=(m_D)_{21},\quad(m_D)_{13}=(m_D)_{23}
\quad \to \quad {\rm 12\,\, patterns}\\
&&(m_D)_{11}=(m_D)_{12}=(m_D)_{21},\quad(m_D)_{22}=(m_D)_{23}
\quad \to \quad {\rm 12\,\, patterns}\\
&&(m_D)_{11}=(m_D)_{12}=(m_D)_{23},\quad(m_D)_{13}=(m_D)_{21}
\quad \to \quad {\rm 12\,\, patterns}\\
&&(m_D)_{11}=(m_D)_{12}=(m_D)_{23},\quad(m_D)_{21}=(m_D)_{22}
\quad \to \quad {\rm 6\,\, patterns}
\label{321ex}
\end{eqnarray}
All 60 textures are generated from the above 6 representatives.

\paragraph{(4,1,1) case} 
There are ${}^6 {\rm C}_4  = 15$
patterns of textures.
The representatives are
\begin{eqnarray}
&&(m_D)_{11} = (m_D)_{12} = (m_D)_{13} = (m_D)_{21} 
\quad \to \quad {\rm 6\,\, patterns}\\
&&(m_D)_{11} = (m_D)_{12} = (m_D)_{21} = (m_D)_{22} 
\quad \to \quad {\rm 3\,\, patterns}\\
&&(m_D)_{11} = (m_D)_{12} = (m_D)_{21} = (m_D)_{23} 
\quad \to \quad {\rm 6\,\, patterns}
\end{eqnarray}
All 15 textures are generated from the above 3 representatives.

\paragraph{(2,2,2) case} 
There are ${}^6 {\rm C}_2 \times {}^4 {\rm C}_2 = 90$
patterns in this category.
It is helpful to remember the case of (2,2,1,1) in 2 equalities.
This case is obtained by imposing equalities between the last
two entries of (2,2,1,1).
As in the case of (2,2,1,1), we should identify the three entries
of (2,2,2).
Then the total number is reduced to $90/3! = 15$ patterns.
The representatives are given by
\begin{eqnarray}
&&(m_D)_{11}=(m_D)_{21},\quad (m_D)_{12}=(m_D)_{22},\quad(m_D)_{13}=(m_D)_{23} 
\quad \to \quad {\rm 1\,\, pattern}\quad\quad\quad\\
&&(m_D)_{11}=(m_D)_{12},\quad (m_D)_{21}=(m_D)_{22},\quad(m_D)_{13}=(m_D)_{23} 
\quad \to \quad {\rm 3\,\, patterns}\quad\quad\\
&&(m_D)_{11}=(m_D)_{22},\quad (m_D)_{12}=(m_D)_{21},\quad(m_D)_{13}=(m_D)_{23} 
\quad \to \quad {\rm 3\,\, patterns}\quad\quad\\
&&(m_D)_{11}=(m_D)_{12},\quad (m_D)_{13}=(m_D)_{21},\quad(m_D)_{22}=(m_D)_{23} 
\quad \to \quad {\rm 6\,\, patterns}\quad\quad\\
&&(m_D)_{11}=(m_D)_{22},\quad (m_D)_{12}=(m_D)_{23},\quad(m_D)_{13}=(m_D)_{21} 
\quad \to  \quad {\rm 2\,\, patterns}\quad\quad\quad
\end{eqnarray}
All 15 textures are generated from the above 5 representatives
by the exchange of the rows and the columns.

\subsection{4 equalities}
\label{4eqmD}
Here we consider 4 equalities in $m_D$.
As in 3 equalities, there are three types of distributions;
(5,1), (4,2) and (3,3).
We study the three cases in turn.

\paragraph{(5,1) case} 
In this case, there are ${}^6 {\rm C}_5 = {}^6 {\rm C}_1 =  6$ 
patterns.
A representative is
\begin{eqnarray}
&&(m_D)_{11}=(m_D)_{12}=(m_D)_{13}=(m_D)_{21}=(m_D)_{22}
\quad \to \quad {\rm 6\,\, patterns}
\end{eqnarray}
All 6 textures are generated from the above representative by 
the exchange of the rows and the columns.

\paragraph{(4,2) case} 
There are ${}^6 {\rm C}_4  = 15$ patterns of textures.
The representatives are
\begin{eqnarray}
&&(m_D)_{11}=(m_D)_{12}=(m_D)_{13}=(m_D)_{21},\quad(m_D)_{22}=(m_D)_{23}
\quad \to \quad {\rm 6\,\, patterns}\quad\quad\quad\\
&&(m_D)_{11}=(m_D)_{12}=(m_D)_{21}=(m_D)_{22},\quad(m_D)_{13}=(m_D)_{23}
\quad \to \quad {\rm 3\,\, patterns}\\
&&(m_D)_{11}=(m_D)_{12}=(m_D)_{21}=(m_D)_{23},\quad(m_D)_{13}=(m_D)_{22}
\quad \to \quad {\rm 6\,\, patterns}
\end{eqnarray}
All 15 textures are generated from the above 3 representatives
by the exchange of the rows and the columns.

\paragraph{(3,3) case} 
There are ${}^6 {\rm C}_3  = 20$ patterns of textures in this case.
However 20 patterns contain redundancy.
We can reproduce all 20 patterns from fundamental 10 patterns
by exchanging the two entries of (3,3).
The 10 patterns can be obtained from the three representatives.
They can be taken as
\begin{eqnarray}
&&(m_D)_{11}=(m_D)_{12}=(m_D)_{13},\quad(m_D)_{21}=(m_D)_{22}=(m_D)_{23}
\quad \to \quad {\rm 1\,\, pattern}\quad\quad\quad\\
&&(m_D)_{11}=(m_D)_{12}=(m_D)_{21},\quad(m_D)_{13}=(m_D)_{22}=(m_D)_{23}
\quad \to \quad {\rm 6\,\, patterns}\\
&&(m_D)_{11}=(m_D)_{12}=(m_D)_{23},\quad(m_D)_{13}=(m_D)_{21}=(m_D)_{22}
\quad \to \quad {\rm 3\,\, patterns}
\label{33ex}
\end{eqnarray}
All 10 textures are generated from the above 3 representatives
by the exchange of the rows and the columns.

\subsection{5 equalities}
\label{5eqmD}
In this case, all the matrix elements in $m_D$ are equal 
and 
the resultant left-handed Majorana mass matrix 
is of democratic form. This provides two massless 
neutrinos together with a nonzero $M_R$.
Thus we can exclude $m_D$ with 5 equalities.


\end{document}